\documentclass[twocolumn,superscriptaddress]{revtex4}%
\usepackage{amssymb}
\usepackage[dvipdfm,colorlinks, citecolor=red]{hyperref}
\usepackage{color}
\usepackage{graphicx}
\usepackage{dcolumn}
\usepackage{bm}
\usepackage[header,title,page,titletoc]{appendix}
\usepackage{amsmath}
\usepackage{amsfonts}%
\providecommand{\U}[1]{\protect \rule{.1in}{.1in}}

\begin{document}
\title{Fermi Arcs in Tilted Weyl Semimetals: Classification, Evolution and Transport Properties}
\author{Xiao-Ming Zhao}
\affiliation{Center for Advanced Quantum Studies, Department of Physics, Beijing Normal
University, Beijing 100875, China}
\author{Xiao Kong}
\affiliation{Center for Advanced Quantum Studies, Department of Physics, Beijing Normal
University, Beijing 100875, China}
\author{Cui-Xian Guo}
\affiliation{Center for Advanced Quantum Studies, Department of Physics, Beijing Normal
University, Beijing 100875, China}
\author{Ya-Jie Wu}
\affiliation{School of Science, Xi'an Technological University, Xi'an 710021, China}
\author{Su-Peng Kou}
\thanks{Corresponding author}
\email{spkou@bnu.edu.cn}
\affiliation{Center for Advanced Quantum Studies, Department of Physics, Beijing Normal
University, Beijing 100875, China}

\begin{abstract}
The Weyl semimetal is a new quantum state of topological semimetal, of which
topological surface states -- the Fermi arcs exist. In this paper, the Fermi
arcs in Weyl semimetals are classified into two classes -- class-1 and class-2.
Based on a tight-binding model, the evolution and transport properties of
class-1/2 Fermi arcs are studied via the tilting strength of the bulk Weyl cones.
The (residual) anomalous Hall conductivity of topological surface states is a
physical consequence of class-1 Fermi arc and thus class-1 Fermi arc becomes a
nontrivial topological property for hybrid or type-II Weyl semimetal.
Therefore, this work provides an intuitive method to learn topological
properties of Weyl semimetal.

\end{abstract}
\maketitle

\section{introduction}
Topological materials, including topological superconductors, topological
superfluids, topological insulators and topological semimetals become more and
more important in condensed matter physics. The Weyl semimetal (WSM) is a new
type of topological semimetals
\cite{BalentsL2011,SavrasovSY2011,NinomiyaM1983,FangZ2011,RanY2011,QiXL2013,VanderbiltD2014,MiyakeT2015,DaiX2015,HuangSM2015}%
: a three dimensional (3D) version of the graphene with several bulk gapless
points in the reciprocal space, i.e., Weyl nodes. Each Weyl node can be
regarded as a monopole in the reciprocal space, carrying a topological charge
of $\pm1$ corresponding to the left-hand or right-hand chirality. The Weyl
nodes are separated in momentum space and exist in pairs which are connected
only through topological surface states -- the Fermi arcs (FAs). The recent
progress in identified WSM materials
\cite{SoljacicM2015,DingH2015,DingHDaiX2015,YanB2015NP,HasanMZ2015} has driven
a flurry of exciting researches to probe the various fascinating phenomena
connected to Weyl fermions, such as the chiral anomaly and the chiral magnetic
effect \cite{NinomiyaN1983,AjiV2012,YamamotoN2012,BurkovAA2012}. The WSMs have
been discovered in the \textrm{TaAs} family where it features multiple Fermi
arcs arising from topological surface states\cite{LvBQ2015,YangLX2015}, and
exhibits novel quantum phenomena, e.g., chiral anomaly induced negative
magnetoresistance \cite{SpivakBZ2013,HuangX2015} and possibly emergent
supersymmetry \cite{YaoH2015}.

Recently, it was proposed theoretically and experimentally that a new type
(type-II) of WSM \cite{SoluyanovAA2015,WangZJ2016,BernevigBA} can emerge with
topologically-protected touching between electron and hole pockets. The
topological surface states were confirmed by directly observing the surface
states using bulk and surface-sensitive angle-resolved photoemission
spectroscopy (ARPES). Besides, the third type of WSM is the hybrid one (or
type-1.5) in which one Weyl node belongs to type-I whereas its chiral partner
belongs to the type-II. The type-II WSM has been proposed to exist in layered
transition metal dichalcogenides, such as W$_{x}$Mo$_{1-x}$Te$_{2}%
$\cite{YanB2015PRB,QiY2016,ChangTR2016}, and the tight-binding models for the
hybrid WSM have been constructed in possible materials and optical lattices
\cite{kongxiao2017,chengang2016}.

It is known that for usual quantum Hall systems, the current-carrying chiral
edge states are responsible for the integer quantized Hall conductance which
can be measured by the transport experiments\cite{Halperin1982,Streda1984}. The
anomalous Hall conductivity (AHC) for a type-I WSM with two Weyl nodes is
proportional to the distance of the Weyl points\cite{RanY2011}. When tilting
the energy dispersion of type-I WSM enough along a certain direction we get a hybrid or type-II WSM, and the Weyl cones could even be tipped over so that the Fermi surface
transforms from a point to a line or a surface. Both the surface states and
bulk states have contribution to the Hall conductance. For type-II WSM the Hall conductivity is not universal and can change
sign as a function of the parameters quantifying the tilting strength in
Ref.\cite{typeII2016}, where they only considered the low energy approximation
of the bulk states. For type-II and hybrid type WSMs, the surface states make significant contribution to the Hall conductance. However, it is not known
how the AHC is related to the surface states for type-II and hybrid type WSMs.
The main purpose of the paper is to make it clear.

This paper is organized as follows. In Sec.\ref{Sec2},  we define the two classes of FAs generally and induce a dimensionless parameter - residual class-1 Fermi arc - to describe the residual effects of the Fermi arc. And then based on a tight-binding model, the evolution of residual Fermi arc with the tilting strength of the Weyl cones is shown in Sec.\ref{sec3}. Furthermore, Sec.\ref{sec4}  shows the effects of residual Fermi arcs on the anomalous Hall conductivity and the restrict relationship between them is figured out. Finally, we conclude our research in Sec.\ref{sec5}.

\section{Classification of the Fermi arcs}\label{Sec2}
A low-energy Hamiltonian of a WSM is written as
\begin{equation}
H_{\mathrm{W}}(k)=\sum_{i}(v_{i}\sigma_{i}+C_{i})k_{i}\label{HW}%
\end{equation}
where $\sigma_{i}$ $(i=x,y,z)$ is the $2\times2$ Pauli matric, $v_{i}$ is
the Fermi velocity, $C_{i}$ denotes the tilting strength of the linear energy
dispersion along $k_{i}$ which breaks Lorentz invariance. The Weyl nodes are
characterized by a topological Chern number $n_{W}=\mathrm{sgn}[\prod_{i}%
v_{i}]=\pm1$. The Weyl nodes have been classified into type-I for the case of
$|C_{i}/v_{i}|<1$ ($i=x,y,z$), and type-II for the case of $|C_{i}/v_{i}|>1$
($i=x,y,z$). \begin{figure}[t]
\scalebox{0.43}{\includegraphics* [0.3in,0.5in][8.5in,5.7in]{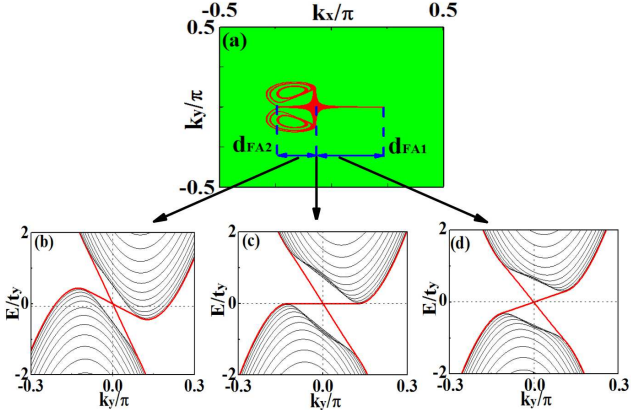}}\caption{(Coloronline)
\textbf{Coexistence of class-1 and class-2 Fermi arcs in hybrid type WSM.}
\textbf{(a)} The two classes of Fermi arcs locate at $k_{y}=0$, which are denoted by blue
arrowed lines. The red arcs denote the shape of electron pockets and hole
pockets. \textbf{(b-d)} Energy dispersion of the surface states (denoted by
red line) for two classes of Fermi arcs.}%
\label{Espectrum}%
\end{figure}

In a WSM, a paired Weyl nodes with opposite chiralities are connected through
the Fermi arc. Generally, when we take open boundary condition (OBC) along one
direction (here take z-direction as an example) the surface states cross Fermi
surface $E_{f}$ somewhere marked by $k_{\lambda}=(k_{\lambda,x},k_{\lambda
,y})$ and eventually the cross points constitute the Fermi arc. The surface
states near a certain point $k_{\lambda}$ can also be described by a linear
model as
\begin{equation}
H_{\mathrm{S}}(k_{\lambda})=(v_{s}\sigma_{z}+C_{\lambda,\perp})\mathbf{k}%
_{\lambda,\perp}%
\end{equation}
where $\mathbf{k}_{\lambda,\perp}$ is the momentum perpendicular to Fermi arc,
$C_{\lambda,\perp}$ represents the tilting strength along the $\mathbf{k}%
_{\lambda,\perp}$ direction. Similar to the classification of Weyl nodes which
is based on the tilting strength of the bulk energy dispersion, we classify
the Fermi arcs into two classes that are denoted by
\begin{align}
\mathrm{Class\text{-}1}\text{ }\mathrm{(FA1)}\text{{}}  &  \text{\textrm{: }%
}|C_{\lambda,{\perp}}|/v_{s}<1,\nonumber \\
\mathrm{Class\text{-}2}\text{ }\mathrm{(FA2)}\text{{}}  &  \text{\textrm{: }%
}|C_{\lambda,{\perp}}|/v_{s}>1.
\end{align}

Because FA1 (FA2) exists near type-I (type-II) Weyl node, in the hybrid WSM,
the FA1 and FA2 may coexist. The schematic diagram of the two classes of Fermi
arcs is shown in Fig.\ref{Espectrum}. When the type-I WSM changes into hybrid
or type-II, the region of FA1 shrinks and the region of FA2 enlarges. To
characterize the evolution of two classes of Fermi arcs, we introduce a
dimensionless parameter, i.e., residual ratio of class-1 Fermi arc
\[
R_{\mathrm{FA1}}=\frac{d_{\mathrm{FA1}}}{d_{\mathrm{0,FA}}}=\frac
{\mathbf{d}_{\mathrm{\mathbf{FA1}}}\cdot \mathbf{e}_{\mathbf{l}}}%
{\mathbf{d}_{\mathrm{\mathbf{0,FA}}}\cdot \mathbf{e}_{\mathbf{l}}}%
\]
where $\mathbf{d}_{\mathrm{\mathbf{FA1}}}$ ( $\mathbf{d}%
_{\mathrm{\mathbf{0,FA}}}$) represents the length-vector of FA1 (FA) from
positive chiral Weyl node to negative chiral Weyl node, and $d_{\mathrm{FA1}}$
($d_{\mathrm{0,FA}}$) is the projection of $\mathbf{d}_{\mathrm{\mathbf{FA1}}%
}$ ( $\mathbf{d}_{\mathrm{\mathbf{0,FA}}}$) on the direction $\mathbf{e}%
_{\mathbf{l}}$ in momentum space. Remember that the length $d_{\mathrm{0,FA}}$
of the FA is constant due to topological protection of the Weyl nodes.
\begin{figure}[t]
\scalebox{0.40}{\includegraphics* [0.0in,0.4in][9.0in,6in]{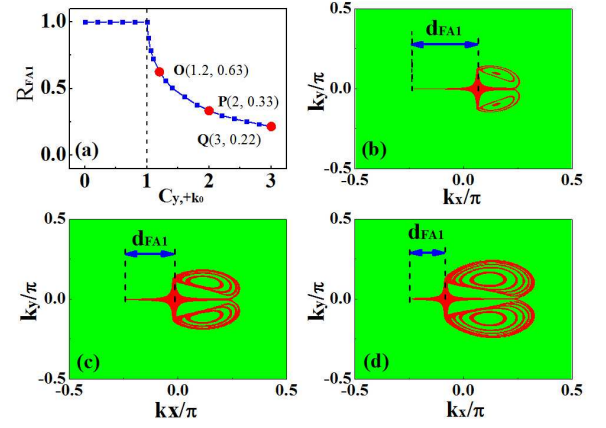}}
\caption{(Coloronline) \textbf{The evolution and appearance of the residual class-1 Fermi arc via tilting strength for case-1.} Case-1: WSM changes from type-I to hybrid type. \textbf{(a)} The evolution of residual ratio of class-1 Fermi arc $R_{\mathrm{FA1}}$. \textbf{(b-d)} The appearance of residual class-1 Fermi arc corresponding to the situation at $\textbf{O},\textbf{P},\textbf{Q}$ respectively, which are marked by red solid dots in panel \textbf{(a)} and the tilting strength are $C_{y,+k_{0}}=1.2,2,3$. The residual FA1 locates at left part of Fermi arc}%
\label{FA1}%
\end{figure}

\section{Evolution of two classes of Fermi arcs in tilted Weyl semimetals}\label{sec3}
    To illustrate the effects of the two classes of
Fermi arcs, and for simplicity but without loss of generality, we consider a
tight-binding model on the cubic lattice, of which the Hamiltonian in momentum
space is given by%
\begin{align}
H(k)  &  =[2t_{x}(\cos k_{x}-\cos k_{0})+m(2-\cos k_{y}-\cos k_{z})]\sigma
_{x}\nonumber \\
&  +2t_{y}\sin k_{y}\sigma_{y}+A_{1}\sin(k_{x}+k_{0})\sin k_{y}\sigma
_{0}\nonumber \\
&  +2t_{z}\sin k_{z}\sigma_{z}+A_{2}\sin(k_{x}-k_{0})\sin k_{y}\sigma_{0},
\end{align}
where $t_{x}$($t_{y}$, $t_{z}$) corresponds to the nearest neighbor hopping
parameter along $x$($y$, $z$)-direction of the lattice. When $m>|t_{x}(1+\cos
k_{0})|$, there are two Weyl points located at $\mathbf{k}=(\pm k_{0},0,0)$.
$A_{1}$ $(A_{2})$ leads to the tilting effect of the Weyl cone at $k_{x}%
=k_{0}$ $(-k_{0})$ along $y$-direction. In this paper, we set $k_{0}=\frac
{\pi}{4}$, $-t_{x}\sin k_{0}=t_{y}=t_{z}=1$. Corresponding to the linear-part of the Weyl node described in Eq.\ref{HW}, we define the tilting strength  near $k_{x}=+k_{0}$, $-k_{0}$ as
\begin{equation}
|C_{y,+k_{0}}|=-\frac{A_{1}\cos(k_{0})}{t_{x}},\text{ \  \ }|C_{y,-k_{0}%
}|=\frac{A_{2}\cos(k_{0})}{t_{x}}%
\end{equation}
respectively. The sign of $C_{y,+k_{0}}$/$C_{y,-k_{0}}$ indicates tilting
direction parallel ($+$) or antiparallel ($-$) to $y$-direction. So when
$|C_{y,+k_{0}}|<1$ and $|C_{y,-k_{0}}|<1$ the WSM is type-I, when
$|C_{y,+k_{0}}|>1$ and $|C_{y,-k_{0}}|>1$ it is type-II, otherwise it is
hybrid type.
\begin{figure}[t]
\scalebox{0.40}{\includegraphics* [0.0in,0.4in][9.0in,6in]{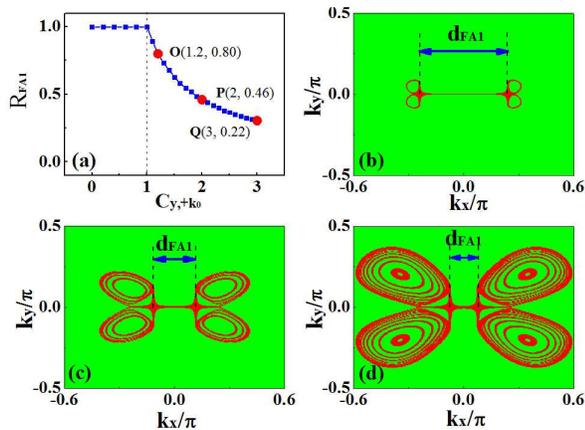}}
\caption{(Coloronline) \textbf{The evolution and appearance of the residual class-1 Fermi arc via tilting strength for case-2.} case-2: WSM changes from type-I to
type-II (node tilting to opposite directions along $k_{y}$ direction). \textbf{(a-d)} Same as panel (a-d) in Fig.\ref{FA1}. The tilting strength are $C_{y,+k_{0}}=1.2,2,3$. The residual FA1 locates at middle part of Fermi arc.}
\label{FA2}
\end{figure}

Based on above model, we study the evolution of two classes of Fermi arcs in tilted
Weyl semimetals. We choose the OBC along $z$-direction and periodic boundary
condition (PBC) along $x/y$-direction. So the momentum-dependence of the
density of states (DOS) is given by
\begin{equation}
\rho(k_{x},k_{y})=\sum_{i_{z}=1}^{N_{z}}-\frac{1}{\pi}\mathrm{Tr}%
[\mathrm{Im}\mathbf{G}(k_{x},k_{y},i_{z})],
\end{equation}
here $\mathbf{G}(k_{x},k_{y},i_{z})=[\mathbf{Z}-\mathbf{H}(k_{x},k_{y},i_{z})]^{-1}$ is the Green's function of the $i_{z}$-th lattice layer along $z$-direction, where $\mathbf{Z}=(E+i\eta)\mathbf{I}$ is the complex energy, $E$ is energy level, $\eta$ is a
infinite small value and $\mathbf{I}$ represent the unit matrix.

\begin{figure}[t]
\scalebox{0.40}{\includegraphics* [0.0in,0.4in][9.0in,6in]{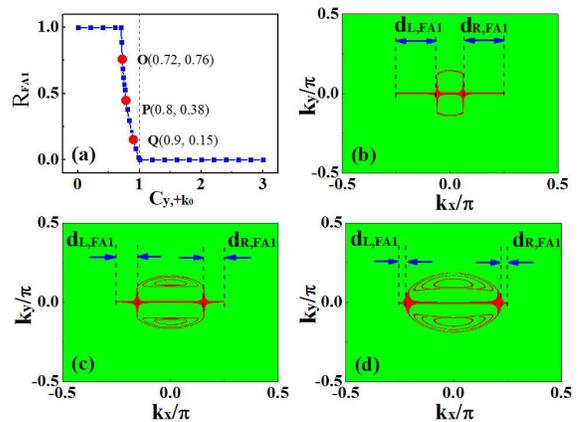}}
\caption{(Coloronline) \textbf{The evolution and appearance of the residual class-1 Fermi arc via tilting strength for case-3.} Case-3: WSM changes from type-I to type-II (node tilting to same directions along $k_{y}$ direction).\textbf{(a-d)} Same as panel (a-d) in Fig.\ref{FA1}. The tilting strength are $C_{y,+k_{0}}=0.72,0.8,0.9$. The residual FA1 have two separated pieces located at both left and right ends.}%
\label{FA3}%
\end{figure}

Fig.\ref{FA1}-\ref{FA3}(\textbf{a}) shows the evolution of residual ratio of class-1 Fermi arc
$R_{\mathrm{FA1}}$ during the evolution of WSM types by tilting the two nodes
toward $y$-direction . We show the results for three
different cases: \textbf{case-1} (Fig.\ref{FA1}): WSM changes from type-I to hybrid type by
tilting the single node at $k_{x}=+k_{0}$ ($C_{y,+k_{0}}>0$, $C_{y,-k_{0}}%
=0$); \textbf{case-2} (Fig.\ref{FA2}): WSM changes from type-I to type-II by tilting the two
nodes along opposite direction ($C_{y,+k_{0}}=C_{y,-k_{0}}$); \textbf{case-3} (Fig.\ref{FA3}):
WSM changes from type-I to type-II by tilting the two nodes along same
direction($C_{y,+k_{0}}=-C_{y,-k_{0}}$). While panels\textbf{ (b-d)} show the features of
the distribution of states on Fermi surface. The residual Fermi arcs in three cases are
marked by red dots $\mathbf{O}$, $\mathbf{P}$, $\mathbf{Q}$ in panel
\textbf{(a)} respectively.

On the one hand, for the type-I WSM, the tilting strength is smaller than the critical value,
i.e., $|C_{y,+k_{0}}|<1$ and $|C_{y,-k_{0}}|<1,$ we have $d_{\mathrm{FA1}}=d_{\mathrm{0,FA1}%
}\equiv2k_{0}$. On the other hand, for the type-II or hybrid type WSM, the tilting strength is larger than
the critical value, i.e., $|C_{y,+k_{0}}|>1$ or $|C_{y,-k_{0}}|>1$, several
pairs of electron-pockets and hole-pockets emerge on the Fermi surface which
are induced by bulk states. For this case, FA1 and FA2 coexist. The residual FA1 locates at left part of Fermi arc
in case-1 and middle part in case-2. In case-3 the residual FA1
have two separated pieces located at both ends, of which the total length of
the residual FA1 is written as $d_{\mathrm{FA1}}=d_{\mathrm{L,FA1}%
}+d_{\mathrm{R,FA1}}$ (see Fig.\ref{FA3}).

\section{Anomalous Hall conductivity induced by Residual class-1 Fermi arc}\label{sec4}
Anomalous Hall effect is a topological property in WSMs, and changing of
topological surface states changes transport properties in WSMs, especially
for AHC. In this part, we explore the relationship between the residual FA1 and AHC for different types of WSMs.

For a WSM, the 3D sample can be divided into 2D slices along a given
direction\cite{BalentsL2011}. In the WSM with two Weyl nodes mentioned above,
the 2D slices for $k_{x}\in(-k_{0},k_{0})$ can be regarded as a 2D topological
insulator and the FA is the set of edge states corresponding to the 2D integer
quantum Hall states. So the AHC of the WSM is proportional to the distance of
the Weyl nodes in the momentum space, i.e., $\sigma_{0,yz}=e^{2}k_{0}/\pi h$.

\begin{figure}[t]
\scalebox{0.38}{\includegraphics* [0.0in,2.3in][9.0in,11.5in]{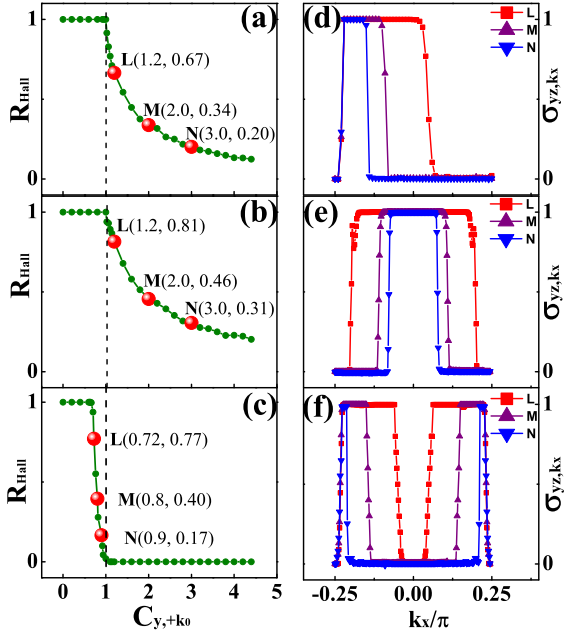}}
\caption{(Coloronline) \textbf{The evolution and momentum-dependence of anomalous Hall conductivity.} \textbf{(a)} The evolution of residual anomalous Hall conductivity $R_{\mathrm{Hall}}$ for the situation of case-1.
\textbf{(d)} The anomalous Hall conductivity structure in $k_{x}$ direction for the situation $\mathbf{L,M,N}$, which are denoted by red solid dots in panel
\textbf{a}. \textbf{(b)(e)} for case-2 and \textbf{(c)(f)} for case-3. }%
\label{HC}%
\end{figure}
Then we discuss the transport properties for the hybrid/type-II WSM. To
characterize the topological properties of transport, we introduce another
dimensionless parameter -- residual ratio of AHC of the tilted WSM,
\begin{equation}
R_{\mathrm{Hall}}=\frac{\sigma_{yz}}{\sigma_{0,yz}}=\frac{\int_{-k_{0}}%
^{k_{0}}\frac{dk_{x}}{2\pi}\sigma_{yz,k_{x}}}{\sigma_{0,yz}},
\end{equation}
where $\sigma_{0,yz}$ is the AHC for type-I WSM and can be used to normalize
the AHC in the tilting process. $\sigma_{yz,k_{x}}$ is the AHC of the 2D
yz-slice for $k_{x}\in(-k_{0},k_{0})$.

\textbf{\emph{Method:}} The method we use to calculate Hall conductivity is Landauer-B\"{u}ttiker theory. The Hall conductivity is described by four-terminal formalism\cite{ButtikerM1986, ShenSQ2005},
\begin{equation}
\sigma_{yz,k_{x}}=\frac{e^{2}}{h}(T_{12}-T_{14}).
\end{equation}
The transmission coefficient $T_{pq}(E_{F})$ from the terminal $p$ to terminal
$q$ is defined by
\begin{equation}
T_{pq}(E_{F})=\mathrm{Tr}[\Gamma^{p}G_{c}^{r}\Gamma^{q}G_{c}^{a}],
\end{equation}
where $p$, $q=1,2,3,4$, and they are ordered clockwise. Terminal-1 and
terminal-3 (terminal-2 and terminal-4) are attached to the surface along the
$+y$ ($+z$)-direction acting as the measurement electrodes. The coupling
matric $\Gamma^{p}$ is determined by the self-energy at the terminal $p$,
$\Gamma^{p}=i[\sum^{p}-(\sum^{p})^{\dagger}$, the self-energy is acted as
\begin{equation}
\sum^{p}=H_{pc}^{\dag}g_{p}^{s}H_{pc},
\end{equation}
where $H_{pc}$ is the coupling matrix between terminal $p$ and center device
area, $g_{p}^{s}$ is the surface Green's function of the lead $p$. $G_{c}^{r}$ ($G_{c}^{a}$) is the
retarded (advanced) Green's function of center device area
\begin{equation}
G_{c}^{r}(E_{F}+i\eta)=[E_{F}+i\eta-H_{c}-\sum \nolimits_{p=1}^{4}\sum
^{p}]^{-1},
\end{equation}
where $H_{c}$ is the real-space Hamiltonian of center device area, and $\eta$
is a infinite small value.
\begin{figure}[t]
\scalebox{0.30}{\includegraphics* [0.0in,0in][12.0in,7.5in]{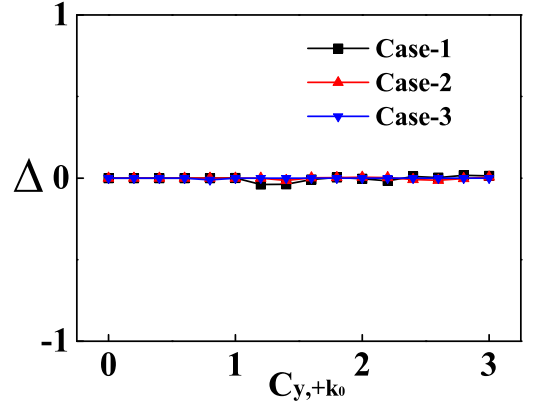}}
\caption{(Coloronline) \textbf{The intrinsic relationship between
$R_{\mathrm{FA1}}$ and $R_{\mathrm{Hall}}$.} Residual Hall conductivity is a
physical consequence of residual class-1 Fermi arc.}%
\label{RfaRhall}%
\end{figure}

To obtain $T_{pq}$ efficiently, we simplify the calculation of $g_{p}^{s}$ and
$G_{c}^{r}$ ($G_{c}^{a}$). On the one hand, in order to determine $g_{p}^{s}$ for these
side-attached leads, the semi-infinite leads are sketched with discrete
effective principal layers. These layers are defined as the smallest group of
neighboring atomic planes and they allow only nearest-neighbor interaction
between them. So we effectively transforms the original system into a linear
chain of principal layers \cite{Sancho1984,TrM2013,Mtd2005,Mtd2006}. Because
the number of iterations required for convergence is smaller than any other
method, this technique is valid \cite{Iterations2008,Iterations2010}. On the
other hand, to calculate $G_{c}^{r}$ for a large real-space system, it
requires a full inversion of the 3D Hamiltonian. In fact, $G_{c}^{r}$ and
$\Gamma^{p}$ is the function of $\sum^{p}$, and the matrix elements of
$\sum^{p}$\ are equal to $0$ except for the elements corresponding to the
surface of center device area which means $\sum^{p}$ is a large sparse matrix.
So we can firstly figure out the Green's function $G_{c,out}^{r}$ and
$\Gamma_{out}^{p}$ which only contain the elements related to outmost layer of
the center device area. It is easy to deduce that%
\begin{equation}
T_{pq}=\mathrm{Tr}[\Gamma_{out}^{p}G_{c,out}^{r}\Gamma_{out}^{q}G_{c,out}%
^{a}],
\end{equation}
and the Green's function of the full device area can be calculated by Dyson
equation $G=G^{0}+G^{0}VG$.

As a result, the transform from 3D full real-space lattice to 2D out-most
layer lattice dramatically improved the  computational efficiency: the processing
times for an initial 3D lattice is scaled as $O[(N_{x}N_{y}N_{z})^{4}]$; For a
2D $yz\mathrm{-slice}$ is $O[N_{x}(N_{y}N_{z})^{4}]$; For the outmost edge of
the $yz\mathrm{-slice}$ is $O[N_{x}N_{y}(N_{z})^{4}]$.

\textbf{\emph{The result and generalization:}} From the results, we found that the residual ratio
of AHC $R_{\mathrm{Hall}}$ via tilting strength of nodes (shown in
Fig.\ref{HC}) are almost the same as the residual ratio of FA1
$R_{\mathrm{FA1}}$ via tilting strength of nodes (shown in Fig.\ref{FA1}-\ref{FA3}). Fig.\ref{HC}\textbf{d} ( or \textbf{e}, \textbf{f}) shows the distribution of
$\sigma_{yz,k_{x}}$ along $k_{x}$-direction for the points $\mathbf{L}$,
$\mathbf{M}$, $\mathbf{N}$ , which are denoted by
red solid dots in panel Fig.\ref{HC}\textbf{a} (or \textbf{b}, \textbf{c}) respectively. When the tilting strength is larger than the critical
value, i.e., $|C_{y,+k_{0}}|>1$ or $|C_{y,-k_{0}}|>1,$ $R_{Hall}$ decreases in
the three cases. In \textbf{case-1} for $C_{y,+k_{0}}>1$, the region of
$\sigma_{yz,k_{x}}=1$ doesn't change near the $k_{x}=-k_{0}$ and with the
increasing of $C_{y,+k_{0}}$ the region of $\sigma_{yz,k_{x}}=1$ shrinks; In
\textbf{case-2} the region of $\sigma_{yz,k_{x}}=1$ pins at the middle of
Fermi arc and the region of $\sigma_{yz,k_{x}}=1$ shrinks with the increasing
of $C_{y,+k_{0}}$; In \textbf{case-3}, the region of $\sigma_{yz,k_{x}}=1$ is
separated into two pieces near the two Weyl points and $R_{Hall}$ begins
to decrease at $C_{y,+k_{0}}\simeq0.7$ and rapidly turns into $0$ before
$C_{y,+k_{0}}=1$. 

Finally we emphasize the intrinsic relationship between the residual ratio of
FA1 $R_{\mathrm{FA1}}$ and the residual ratio of AHC
$R_{\mathrm{Hall}}$, i.e.,
\begin{equation}
R_{\mathrm{Hall}}\simeq R_{\mathtt{FA1}}.\label{eq}%
\end{equation}
and the difference of the two dimensionless parameters $\Delta$ is figured in
Fig.\ref{RfaRhall}. This result indicates that the residual Hall conductance
comes from the contribution of the residual FA1.

It is known that the Hall conductivity is a physical consequence of Fermi arc
and becomes a nontrivial topological property for type-I WSM. The intrinsic
relationship between the residual ratio of FA1 and the residual ratio of AHC
indicates that the residual Hall conductivity is a physical consequence of
residual FA1 and also becomes a nontrivial topological property for hybrid and
type-II WSM.

In general, a material may have multiple pairs of nodes and different nodes
in the same system may belong to different types. For example, as the
inversion symmetry-breaking material, the TaAs family belongs to type-I WSM
which contains $24$ bulk Weyl cones, and WTe$_{2}$ belongs to type-II WSM
which contains eight Weyl cones. And in the systems that don't have both
inversion and time-reversal symmetries, one may find hybrid WSM with mixed
types of Weyl nodes. Similar to the WSM with a pair of Weyl nodes, we may
generalize our approach to those complicated systems and define several
residual ratios of FA1/AHC as
\begin{equation}
\widetilde{R}_{\mathtt{FA1}}=\sum_{i}\frac{\mathbf{d}_{\mathtt{FA1}}%
\cdot \mathbf{e}_{k}}{d_{0,i}},\text{ \ }\widetilde{R}_{Hall}=\sum_{i}%
\frac{\mathbf{\sigma}_{i}}{\sigma_{0,i}},
\end{equation}
where we sum over the length of all pieces of Fermi arcs in the system along
$\mathbf{e}_{k}$-direction in momentum space. Similarly, the intrinsic
relationship between $\widetilde{R}_{arc}$ and $\widetilde{R}_{Hall}$ is
similar to Eq.\ref{eq}.

\section{Conclusion and discussion} \label{sec5}
In conclusion, we find that there exist
two classes of Fermi arcs in a Weyl semimetal. Based on a tight-binding model,
the evolution and transport properties of the two classes of Fermi arcs are
studied. The results show that the FA1 shrinks with the
increasing of the tilting strength of the Weyl cones, which results to the
decrease of anomalous Hall conductance. To characterize the interplay effect
between the FA1 and AHC, we introduce two
dimensionless parameters -- the residual ratio of FA1
$R_{\mathrm{FA1}}$ and the residual ratio of AHC
$R_{\mathrm{Hall}}$. In particular, there exists an intrinsic relationship
between $R_{\mathtt{FA1}}$ and $R_{Hall}$, i.e., $R_{\mathtt{FA1}}\simeq
R_{Hall}.$ This result indicates that the residual Hall conductivity is a
physical consequence of residual class-1 Fermi arc and that becomes a nontrivial
topological property for hybrid or type-II Weyl semimetal. Therefore, this
work provides an intuitive method to learn topological properties of Weyl semimetals.

\acknowledgments
This work is supported by NSFC under Grants No. 11474025,11504285,11674026,
SRFDP, and the Young Talent fund of the University Association for Science and
Technology in Shaanxi, China.

\end{document}